\documentclass[letter,twocolumn,10pt]{article}
\usepackage{procIAGssymp}   
\usepackage{epsfig}

\newcommand{\ntab}[2]{ \multicolumn{1}{#1}{#2} }
\newcommand{\beq}{\begin{equation}}
\newcommand{\eeqn}{\nonumber\end{equation}}
\newcommand{\dss}{\displaystyle}
\newcommand{\eeq}[1]{\label{#1}\end{equation}}
\newcommand{\Frac}[2]{\frac{\displaystyle\strut #1}{\displaystyle\strut #2} }

\title{Atmospheric pressure loading for routine data analysis}
\author{Leonid Petrov 
\thanks{NVI/NASA Goddard Space Flight Center, Greenbelt, Maryland, USA (Leonid.Petrov@gsfc.nasa.gov)}
\and    Jean-Paul Boy 
\thanks{NASA Goddard Space Flight Center, Greenbelt, Maryland, USA (boy@bowie.gsfc.nasa.gov)} }
\date{ }
\begin{document}
\maketitle 

\paragraph{Abstract.}
We have computed 3-D displacements induced by atmospheric pressure 
loading from 6-hourly surface pressure field from NCEP (National Center for 
Environmental Predictions) Reanalysis data for all Very Long Baseline  
Interferometry) and SLR (Satellite Laser Ranging) stations. We have 
quantitatively estimated the error budget our time series of pressure loading
and found that the errors are below 15\%. We validated our loading series
by comparing them with a dataset of 3.5~million VLBI observations for the 
period of 1980--2003. We have shown that the amount of power which is present 
in the loading time series, but does not present in the VLBI data is, 
on average, only 5\%. We have also succeeded, for the first time, to detect
horizontal displacements caused by atmospheric loading. 
The correction of atmospheric loading in VLBI data allows a significant 
reduction of baseline repeatability, except for the annual component.

\section{Introduction}

Atmospheric pressure variations which can reach 50~mbar induce deformation 
of the solid Earth at the level of several centimeters, which certainly
should be taking into account in routine analysis of space geodesy 
observations. Atmospheric loading effects are computed by convolving the 
surface pressure field with Green's functions (see, for example, 
Farrell (1972)). 

van Dam and Herring (1994) and van Dam~et~al. (1994) have 
computed  atmospheric loading using a former atmospheric model from 
the NMC (National Meteorological Center) with $2.\!^\circ5$~by~$2.\!^\circ5$ 
spatial resolution and 12~hour sampling and applied to VLBI (Very Long 
Baseline Interferometry) and GPS (Global Positioning System) 
site position data. Analysis of reduction of variance of baseline lengths 
showed that atmospheric loading signal was clearly present in VLBI and GPS 
datasets. However, only 62\% of the power of the signal computed using their
model was found in the VLBI data. This unexplained discrepancy and very high
computational cost of pressure loading calculation are the reasons why
modeling atmospheric pressure loading computed using the global numerical
weather model did not come into practice of routine data analysis.

%

There are several reasons which motivated us to re-visit this topic:
the accuracy of VLBI geodetic observations has increased during the last ten 
years which has improved our ability to detect tiny crustal motions. The 
NCEP/NCAR (National Center for Environmental Predictions~/ 
National Center for Atmospheric Research) Reanalysis project (Kalnay~et~al. 
(1996)) provides us a continuous and uniform dataset of surface pressure 
with a 6~hour temporal resolution on a $2.\!^\circ5$~by~$2.\!^\circ5$ grid for
the period 1948--now. Because of the computer and network improvements, 
it is now possible to compute on a operational basis the series of the 
atmospheric loading and apply them in a routine data analysis process, for 
example, for the Earth orientation service.

In section~2, we describe our model of the atmospheric pressure loading 
and present the error budget of the loading time series. In section~3, 
we validate the time series of site displacements induced by pressure loading 
by re-analyzing a dataset of 3.5~million of VLBI observations for the period 
1980--2003. We have estimated global admittance factors for vertical and 
horizontal components. We also computed the reduction of variance of baseline 
lengths in order to compare our results with van Dam and Herring (1994). 
Discussion and concluding remarks are given in section~5. Brief outlines 
of the atmospheric pressure loading service are given in section~6.

\section{Computation of atmospheric pressure loading}

\subsection{Characteristics of atmospheric pressure loading}

Displacements at the Earth's surface induced by surface pressure loading
are computed by convolving Green's functions (Farrell (1972)) with the
surface pressure field from NCEP Reanalysis (Kalnay~et~al. (1996)).

We model the ocean response to pressure forcing as the inverted barometer (IB).
It has been shown that this model is adequate for describing sea height 
variations for periods typically longer than 15--20~days (see, for example, 
Wunsch and Stammer (1997)). However the ocean response to atmospheric pressure 
forcing significantly deviates from the IB hypothesis at higher frequencies 
(Tierney~et~al. (2000)). We also assume that enclosed and semi-enclosed seas 
respond to atmospheric pressure as a "non-inverted barometer", i.e. pressure 
variations are fully transmitted to sea floor.

\begin{figure}[!h]
\centerline{
   \epsfig{file=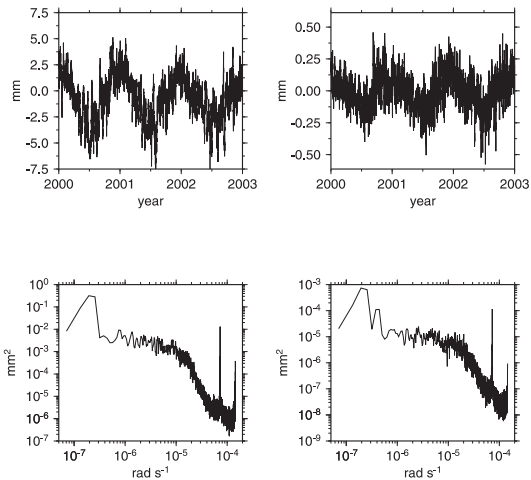,width=2.75in} }
   \caption{Vertical (left) and north (right) displacement induced by atmospheric loading
            for Hartrao station.}
\label{f:hart}
\end{figure}

\begin{figure}[!h]
\centerline{
   \epsfig{file=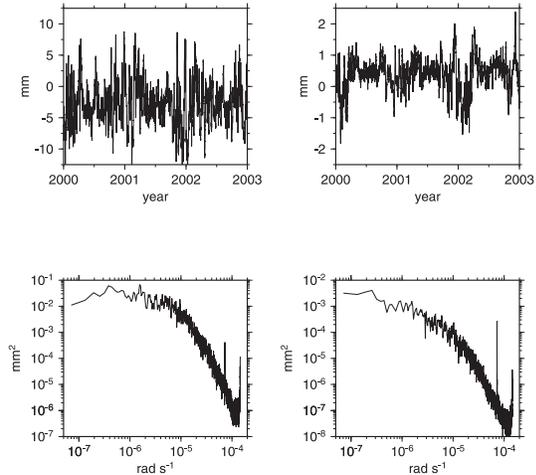,width=2.75in}  }
   \caption{Vertical (left) and east (right) displacement induced by atmospheric loading
            for Wettzell station.}
\label{f:wett}
\end{figure}

Figure~\ref{f:hart} and~\ref{f:wett} shows the time series and their power 
spectrum for the period 2000--2003 at the Hartrao and Wettzell stations. 
They are representative of mid-latitude and equatorial inland stations.

All station displacements show significant narrow-band diurnal ($S_1$), 
semidiurnal ($S_2$) and annual ($Sa$) signals. Displacements for low-latitude 
stations are characterized by a strong wide-band annual and semi-annual signal
and relatively weak amplitudes for periods below 10~days 
(except the $S_1$ and $S_2$ peaks).

Mid latitude stations show the opposite behavior; the mid-latitude atmospheric 
circulation is determined by the circulation of low and high pressure structures 
with typical period of 5--10~days. These periods are also the limit of the 
validity of the IB assumption.

\subsection{Error budget}

We would like to quantitatively asses the error budget of our atmospheric 
loading series. We identified four major sources of errors: 
1) errors in the Green's functions, 2) errors in the surface pressure field, 
3) errors in the land-sea mask and, 4) mismodeling of the ocean response 
to pressure forcing.

The Green's functions are computed for a spherically symmetric, non rotating, 
elastic and isotropic (SNREI) Earth model, adopting PREM (Dziewonski and 
Anderson (1981)) elastic parameters. Therefore, we neglect the effects induced 
by Earth's anelasticity and ellipticity. The differences between our Green's 
functions and Green's functions for an anelastic Earth model (see, for example, 
Pagiatakis (1990)) are typically below 1-2\%. The effect induced by the 
Earth's ellipticity is of the order of magnitude of the Earth's flattening, 
i.e. 0.3\%.

Another source of possible errors is the errors in the surface pressure field 
from NCEP Reanalysis. The first way to estimate them is to compare NCEP 
pressure field with ground pressure measurements. Velicogna~et~al. (2001) 
compared the differences between this atmospheric model and direct barometric 
observations. We would like to quantitatively asses the error budget, but 
since the spatial coverage of barometric measurements is not homogeneous, 
we choose another way to characterize possible errors, and we estimated the 
differences between two NCEP numerical weather models: NCEP Reanalysis and 
NCEP Operational Final Analysis. We have computed the station displacements 
with the NCEP Operational surface pressure field with a spatial resolution 
of $1.\!^\circ0$. The RMS of the differences between the 3-D displacements 
computed with both pressure fields is, on average, about 10\%. However, 
these differences are larger in mountainous areas; this is due to the fact 
that the $2.\!^\circ5$, or even the $1.\!^\circ0$ spatial resolution of the
NCEP Reanalysis and NCEP Operational datasets is not sufficient to model the 
topography in mountainous areas and, therefore, the surface pressure variations.

Since the $2.\!^\circ5$ resolution of the NCEP Reanalysis data is not 
sufficient to correctly represent the coastline, we choose a land-sea mask with 
a $0.\!^\circ25$ resolution derived from FES99 (Lef\`{e}vre~et~al. (2002)) 
ocean tidal model. In order to estimate the errors induced by the land-sea 
mask, we have computed 3-D site displacements with two land-sea masks with 
$0.\!^\circ25$ and $0.\!^\circ50$ resolution. The differences between the two 
time series do not exceed 5\%. However the differences between the loading 
estimates with the $2.\!^\circ5$ and $0.\!^\circ25$ land-sea masks can reach 
10\% for the vertical components and for the 30\% horizontal components, even
for an inland station like Wettzell (Germany) 500~km far from the coasts.

In order to estimate the errors induced by the mismodeling of the ocean 
response to pressure at high frequency, we compared ocean bottom pressure 
variations, as well as the induced loading effects, from two runs of the CLIO 
(Coupled Large-scale Ice Ocean) global circulation model (Goosse and 
Fichelet (1999)). The first run is forced by atmospheric pressure, surface 
winds and heat-fluxes (de Viron~et~al. (2002)), and the other one is forced 
only by winds and heat-fluxes, i.e. assuming an IB response. The differences 
between these two runs are, therefore, interpreted as the departure of the 
ocean response to pressure forcing to the IB hypothesis. We have found that 
the mean vertical and horizontal differences are respectively below 
10\% and 20\%. As expected these differences can be larger for island 
stations or stations close to the coast.

\begin{table}[!h]
   \caption{Global error budget of atmospheric loading estimates.}
   \vspace{0.2cm}
\centerline{
   \small
   \begin{tabular}{l @{\hspace{5em}} r}
       \hline
          Error source            &         rms       \\
       \hline
          Green's functions       &          2~\%     \\
          Surface pressure field  &         10~\%     \\
          Land-sea mask           &          5~\%     \\
          Ocean response          &         10~\%     \\
       \hline
          Total                   &         15~\%     \\
       \hline
   \end{tabular} }
   \label{t:error}
\end{table}

Table~\ref{t:error} gives a summary of the global error budget. Combining 
all sources of errors, we can evaluate the uncertainty of the site 
displacements induced by atmospheric pressure loading at 15\%.

\section{VLBI data processing}

\subsection{VLBI data processing}

The processing of VLBI observations of time delays mostly follows the 
procedure described and Sovers~et~al (1998) and outlined in the IERS
Conventions (McCarthy, 1996). Semi-empirical IERS96 nutation expansion 
was used. We modeled ocean tidal loading with GOT00 (Ray (1999)) model 
for diurnal and semi-diurnal waves, NAO99 (Matsumoto~et~al. (2000)) model 
for long-period zonal waves, and equilibrium tide for the polar tides and 
18.6~year wave. NMF Mapping functions developed by Niell (1996) were used for 
modeling tropospheric path delay.

\subsection{Analysis of global admittance factor}

In solution~G1 we computed admittance factors of the vertical, east and 
north components of the atmospheric pressure loading time series 
displacements into VLBI time delay, averaged over all sites. 

These admittance factors entered the estimation model the following way:

\beq
   \tau = \dss \sum_{i=1}^3 
                 A_{i1} a_{i1} \Frac{\partial \tau}{ \partial r_{i1}} -
                 A_{i2} a_{i2} \Frac{\partial \tau}{ \partial r_{i2}} + \ldots
\eeq{e:e1}
where $\tau$ is VLBI time delay, $a_{ij}$ is the topocentric vector of the 
modeled site displacements due to atmospheric pressure loading for the $j$th 
station, $r_{ij}$ is the topocentric site coordinate vector. The index $i$ runs
through up, east and north components of the topocentric radius-vector. The 
unknown vector of the admittance factors $A$ was estimated in a single 
weighted LSQ fit together with site position and velocities, source 
coordinates, Earth orientation parameters (EOP) as well as 1.2 million nuisance 
parameters.

It can be shown that these admittance factors can be interpreted as 
correlation coefficients between the true (unknown) site displacements and 
our loading estimates within the validity of the assumption that the time 
series of the modeled site position displacements caused by atmospheric 
pressure loading and unmodeled contribution to time delay are 
{\em not correlated}. We refer the reader to Petrov and Boy (2003) for more 
details.

We performed 4 different solutions. In solution G1 we treated the 
three-dimensional vector of the admittance factor as a global parameter, 
which is common for all sites. Then we passed the time series of the site 
position variations through a narrow-band filter and extracted the annual 
component of the pressure loading signal. In the solution G2 we estimated 
admittance factors for these time series. In the solution G3 we estimated 
the admittance factor to the atmospheric pressure loading series with the 
annual component filtered out. Results are presented in Table~\ref{t:gloadm}. 
We also performed another solution in which we estimated admittance factors 
for each individual station. 

The estimates of the admittance factors of the time series of atmospheric
pressure loading with removed annual component are closer to unity.
Annual signals in site positions may be induced by other unmodeled signals 
such as hydrologic loading or thermal antenna height deformation 
(Nothnagel~et~al. (1995)) which are not modeled in this study. 
For example, van Dam~et~al. (2001) showed, that the contribution of 
continental water loading can reach several cm at seasonal time scales. 

Contrary to wide-band signals, independent narrow-band signals are almost
always correlated. Thus, condition of validity of this test is violated when
we consider the annual signal.

\begin{table}[!h]
   \caption{Global Admittance Factors from VLBI solutions.}
   \vspace{0.2cm}
   \small
\centerline{
   \begin{tabular}{l @{\hspace{0.5em}} r @{\hspace{1.5em}} r @{\hspace{1.5em}} r}
       \hline
          Solution & \ntab{c}{Up} & \ntab{c}{East} & \ntab{c}{North} \\
       \hline
         G1 & 0.95 $\pm$ 0.02  &   1.16 $\pm$ 0.06     &  0.84 $\pm$ 0.07    \\
         G2 & 0.46 $\pm$ 0.09  &   1.08 $\pm$ 0.26     & -0.89 $\pm$ 0.26    \\
         G3 & 0.98 $\pm$ 0.02  &   1.21 $\pm$ 0.06     &  1.02 $\pm$ 0.07    \\
       \hline
   \end{tabular} }
   \label{t:gloadm}
\end{table}


Although the mean admittance factors are very close to unity, the admittance 
factors for individual stations are not always close to unity. We can often 
link anomalous admittance with the site positions. Stations located in 
mountainous areas or in the vicinity of the oceans are usually characterized 
by low admittance factors, even sometimes negative values. This can be 
explained, as we showed in section~2.2, by errors in the atmospheric pressure 
field on mountainous areas and errors induced by the mismodeling of oceanic 
response to atmospheric pressure loading.

\subsection{Analysis of reduction of variance coefficients}

  In order to compare our results with results of van Dam and Herring (1994), 
we made another three solutions; in solution~B1, we did not apply 
the atmospheric loading contribution, but we did it in the solution~B2. 
In solution~B3 we have applied the contribution of the atmospheric pressure
loading time series with the annual component filtering out. 

  We used the reduction of variance coefficient~$R$ of baseline lengths as
a measure of agreement of the atmosphere pressure loading displacements series
with observations:
\beq
 R = \frac{\Delta \sigma^2 + \sigma_m^2}{2 \sigma_m^2}
\eeq{e:variance}
where $\Delta \sigma^2$ is the difference between the mean square of 
baseline length residuals before and after adding the contribution due to
station displacements caused by the atmospheric pressure loading, 
and $\sigma_m$ is the variance of the atmospheric loading.

  A linear model was fitted in the series with discontinuities at epochs 
of seismic events for several stations. The weighted root mean square 
of residual baseline lengths was computed for all baselines with more than 
100 sessions for B1, B2 and B3 solutions. 69 baselines fitted this 
criterion. The coefficients of reduction of variance were computed using 
baseline length variances. The mean coefficient of reduction of variance of 
the B2 solution with respect to the reference solution B1 is 
\mbox{ $0.86 \pm 0.04$ }. The mean coefficient of reduction of variance of 
the B3 solution with respect to the reference solution B1 is 
\mbox{ $0.92 \pm 0.04$ }. 

  Van Dam and Herring (1994) found a reduction of variance of $R = 0.62$ and 
$R = 0.76$ without the annual component. Our estimates are significantly
closer to unity than theirs. We attribute these difference partly to 
improvements in modeling atmospheric loading, and partly to improvements 
of geodetic observations. Van Dam and Herring used an older atmospheric model 
provided by NMC (National Meteorological Center) with the same spatial 
resolution and a 12~hour temporal sampling compared to the 6-hours for the 
NCEP Reanalysis model. The NCEP Reanalysis is also a uniform and continuous 
dataset, whereas the NMC pressure field had several discontinuities related 
to changes in the model. We used a land-sea mask with a higher resolution 
($0.\!^\circ25$) for modeling the IB response. 

\section{Discussion and Conclusion}

  A priori estimates of the errors of our time series of the site displacements 
induced by atmospheric pressure loading are less than 15\%. The estimates
of the admittance factors from  the solution G1 demonstrate that on average
95\% of the power of the modeled pressure loading signal presents in the data
well within the error budget. It allows us to make an important conclusion
that in average our model of the atmospheric pressure loading 
{\em quantitatively} agrees with observations. The existence of the significant
discrepancy between the model of the displacements caused by the atmospheric 
pressure loading and observations reported by van Dam and Herring (1994) has
not been confirmed. Except for the annual component, applying the atmospheric 
loading model results in a reduction of the power of the residual harmonic 
site position variations. For the first time we have detected horizontal 
component of atmospheric pressure loading in VLBI observations. 

  Although we have an excellent agreement between the model of atmospheric
pressure loading and the observations in average, the estimates of the 
admittance factors significantly deviate from unity for some individual 
stations, suggesting deficiency of the model for these sites. These stations 
are located either close to the coastline or in mountainous regions. In the 
first case, improvement of atmospheric loading modeling requires to model the 
high-frequency ocean response to pressure forcing which significantly deviates 
from the IB assumption. In the other case, the spatial resolution of NCEP 
Reanalysis data ($2.\!^\circ5$) is too coarse to model the topography and 
therefore atmospheric pressure variations.

We evaluate the effects of modeling atmospheric loading  on EOP estimation. 
We made two solutions, one with applying the contributions of the 3-D 
atmospheric loading displacements, the other without. The RMS differences 
in polar motion and the UT1 angle between these two solutions are typically 
about 100~prad, which is 2--4 smaller than the current EOP uncertainties. 

\section{Service of the atmospheric pressure loading displacements}

  The tests described above allowed us to make a conclusion that the model
of the atmospheric pressure loading has passed validation tests. We have 
computed continuous atmospheric pressure loading series for all VLBI and 
SLR sites starting from 1976.05.05. We found that when the atmospheric 
pressure loading series are computed for two sites separated by $0.\!^\circ05$, 
the differences in site displacements never exceed the 1\% level. Therefore, 
in the case if several VLBI or SLR stations are located within 3 km from each 
other, the atmospheric pressure loading was computed only for one station and 
considered common for all stations located within this site. 

  The first epoch of the atmospheric pressure loading time series is three 
days before the first observation used in data analysis and the last epoch 
is three days after the last observation of the site in the case if all 
stations at the site has ceased operations. Thus, these epochs are different 
for each site. 

  The series of the atmospheric pressure loading for the sites, which are 
considered as active, i.e. continuing observations, are updated every day. 
The file with surface pressure from the NCEP Reanalysis for the last year 
is retrieved by ftp, split into monthly sections and stored. 
The atmospheric pressure loading time series for active sites are augmented 
if the surface pressure files contain data for the epochs for which the 
pressure loading has not been computed. The NCEP Reanalysis numerical weather 
model are updated daily with the time lag 3--7 days. Our atmospheric pressure 
loading time series also updated with the time lag 3--7 days.

  Starting from December 2002 the atmospheric pressure loading 
contribution is incorporated in the model of VLBI data reduction in all 
solutions of the Goddard VLBI group, including operational EOP solutions.
The atmospheric pressure loading time series are available on the Web 
at \texttt{http://gemini.gsfc.nasa.gov/aplo/} for all everybody without 
restrictions. 

\section*{References}
\parindent=0pt
\def \cita{\par \hangindent=4mm \hangafter=1}

\cita de Viron, O., H. Goosse, C. Bizouard and S. Lambert (2002). High-frequency non-tidal
effect of the ocean on the Earth's rotation, {\it EGS 27th General Assembly}. Nice, France.

\cita Dziewonski, A.M. and D.L. Anderson (1981).  Preliminary Reference Earth Model. {\it
Phys. Earth Planet. Inter.}, 25, pp.~297--356.

\cita Farrell, W.E. (1972). Deformation of the Earth by surface loads. {\it Rev. Geophys. Space
Phys.}, 10, pp.~751--797.

\cita Goosse, H. and T. Fichelet (1999). Importance of ice-ocean interactions for the ocean
circulation: a model study. {\it J. Geophys. Res.}, 104, pp.~23,337--23,355.

\cita Kalnay, E., M.~Kanamitsu, R.~Kistler, W.~Collins, D.~Deaven, 
L.~Gandin, M.~Iredell, S.~Saha, G.~White, J.~Woollen, Y.~Zhu, 
A.~Leetma, R.~Reynolds, M.~Chelliah, W.~Ebisuzaki, W.Higgins, J.~Janowiak, 
K.~C.~Mo, C.~Ropelewski, J.~Wang, R.~Jenne amd D.~Joseph (1996). The NCEP/NCAR 
40-Year Reanalysis Project. {\it Bull. Am. Meteorol. Soc.}, 77, pp.~437--471.

\cita Kalnay et al., (1996). The NCEP/NCAR Reanalysis Project. 
{\it Bull. Am. Meteorol. Soc.}, 77, pp.~437--471.

\cita Lef\`{e}vre, F., F.H. Lyard, C. Le Provost and E.J.O. Schrama (2002). FES99: a global tide
finite element solution assimilating tide gauge and altimetric information. {\it J. Atmos.
Oceanic Technol.}, 19, pp.~1345--1356.

\cita MacMillan, D.S. and J.M. Gipson (1994). Atmospheric pressure loading parameters from very
long baseline interferometric observations. {\it J. Geophys. Res.}, 99, pp.~18,081--18,087.

\cita Matsumoto, K., T. Takanezawa, and M. Ooe (2000). Ocean Tide Models Developed by 
Assimilating TOPEX/POSEIDON Altimeter Data into Hydrodynamical Model: A Global 
Model and a Regional Model Around Japan. {\it J. of Oceanog.}, 56, pp.~567--581

\cita McCarthy, D.D. (1996). IERS Convention, {\it IERS Technical Note}, 21, Paris.

\cita Niell, A.E. (1996). Global mapping functions for the atmosphere delay at
radio wavelengths. {\it J. Geophys. Res.}, 100, pp.~3227--3246.

\cita Nothnagel,~A., M.~Pilhatsch, and R.~Haas (1995). Investigations of thermal height 
changes of geodetic VLBI radio telescopes. {\it Proceedings of the 10th 
Working Meeting on European VLBI for Geodesy and Astrometry}, edited by 
R.~Lanotte and G.~Bianco, Agenzia Spatiale Italiana, Matera, pp~.121--133.

\cita Pagiatakis, S.D. (1990). The response of a realistic Earth to ocean tide loading. {\it Geophys.
J. Int.}, 105, pp.~541--560.


\cita Petrov, L. and J.-P. Boy (2003). Study of the atmospheric pressure loading signals in VLBI
observations. Submitted to {\it J. Geophys. Res.}

\cita Ray, R.D. (1999). A global ocean tide model from TOPEX/POSEIDON Altimetry: 
GOT99.2. {\it NASA/TM-1999-209478}, Greenbelt, USA.

\cita Sovers,~O.J, J.L. Fanselow, and C.S. Jacobs (1998). Astrometry and geodesy with
radio interferometry: experiments, models, results. {\it Reviews of Modern 
Physics}, 70, pp.~1393--1454.

\cita Tierney, C., J.M. Wahr, F. Bryan and V. Zlotnicki (2000). Short-period oceanic circulation:
implications for satellite altimetry. {\it Geophys. Res. Lett.}, 27, pp.~1255--1258.

\cita van Dam,~T.M., J.~Wahr, P.C.D.~Millly, A.B.~Shmakin, G.~Blewitt,
D.~Lavallee, and K.M.~Larson (2001). Crustal displacements due to continental water 
loading. {\it Geophys. Res. Lett.}, 28, pp.~651--654.

\cita van Dam, T.M. and T.A. Herring (1994). Detection of atmospheric pressure loading using
Very Long Baseline Interferometry measurements. {\it J. Geophys. Res.}, 99, pp.~4505--4518.

\cita van Dam, T.M., G. Blewitt and M. Heflin (1994). Detection of atmospheric pressure loading
using the Global Positioning System. {\it J. Geophys. Res.}, 99, pp.~29,939--29,950.

\cita Velicogna, I., J.M. Wahr and H. Van den Dool (2001). Can surface pressure be used to removed
atmospheric contributions from GRACE data with sufficient accuracy to recover hydrologic signals?
{\it J. Geophys. Res.}, 106, pp.~16,415--16,434.

\cita Wunsch, C. and D. Stammer (1997). Atmospheric loading and the "inverted barometer" effect.
{\it Rev. Geophys.}, 35, pp.~117--135.

\end{document}